# Mode analysis of Talbot effect with simplified modal method

Shubin Li[1]* and Yancong Lu[2]

[1]Singapore-MIT Alliance and Research and Technology Centre, 138602, Singapore

[2]Shanghai Institute of Optics and Fine Mechanics, Chinese Academy of Sciences

Shanghai 201800, P. R. China

Corresponding author: shubin@smart.mit.edu

**Abstract**

We report the first observation of the periodical properties for Talbot effect with π phase jump. Analytical expressions are derived from simplified modal method to analyze the novelty phenomenon of the Talbot effect with π phase jump, which can deepen our understanding of physical diffraction process. Importantly, the physical reason of π phase jump can be attributed to that the two even grating modes make the left derivative and right derivative of real part of the $E_1$ opposite in sign, which results in the physical information of first order diffractive wave hidden in the near field Talbot effect image. We expect that this theoretical work will be helpful for the tremendous potential applications of the Talbot effect.

## Introduction

The Talbot effect [1], a classical optical phenomenon, has attracted considerable interest because of wide applications ranging from optical test and metrology [2], image process and photolithography, Talbot array illumination [3], and nanostructure fabrication. Self-imaging is also demonstrated for high density grating with period comparable to the incident wavelength.

For simulating the Talbot effect of a high density grating, the near-field image should be calculated by the rigorous couple wave analysis [4]. Unfortunately, this pure numerical method cannot give much physical insight. It should be noted that the simplified modal method, a physical view, has been applied to interpret the diffraction process for rectangular grating [5], triangular grating [6], and slanted grating [7]. In recent study [8], the simplified modal method was firstly introduced to explain the polarization-independent Talbot effect for a high density grating with physical insight. Now, new advances have been obtained to interpret the Talbot effect, which can deepen the understanding of the key nature of Talbot effect.

In this letter, the mode analysis of Talbot effect with $\pi$ phase jump is presented. The diffraction efficiencies and phase can periodically reappear, and period $T$ can be explained by the simplified modal method. The physical reason of $\pi$ phase jump can be attributed to that the two even grating modes make the left derivative and right derivative of real part of the $E_1$ opposite in sign, which can be analyzed by the a series of analytic formulas derived from simplified modal method. The reason of hidden physical information of first order diffractive wave is that $\pi$ phase jump leads to that the electric field $E_1$ is zero. Ultimately, we expect that this new results can provide a new degree of freedom for controlling diffraction patterns via Talbot effect, for applications such as nanostructure fabrication, biological optics, etc..

**Modal method analysis**

Figure 1 shows the schematic diagram of a simulation structure with a simple rectangular grating. A TE polarization plane wave is normal incident on the grating with the wavelength $\lambda$=1550nm. $n_1$=1 and $n_2$=1 are the refractive indices of top and bottom substrate, respectively.

Also, $n_g$ =1.44462 and $n_r$=1 are the refractive indices of grating materials in grating region. $d$ is the period and $h$ is the grating depth. $b$ is the width of the grating ridge and $f=b/d$ is the duty cycle.

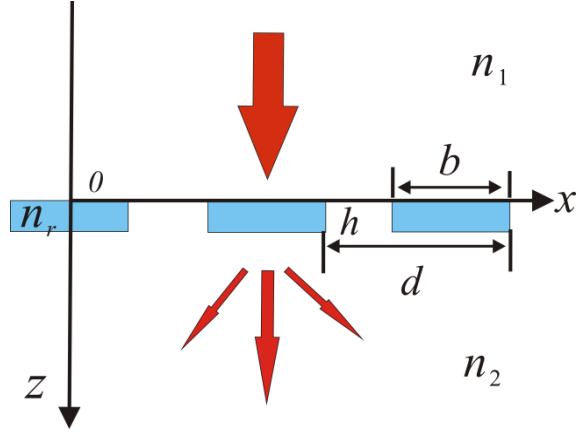

Figure 1 Grating geometry

For a high density grating, only a few diffractive orders can propagate since other orders are evanescent waves. For simplicity, we will limit the period in the range of $\lambda \sim 2\lambda$. Thus, we can just calculate three diffractive orders to reconstruct the near field image [8]:

$$E_y = \hat{y}\sum_m E_m \exp(i\varphi_m)\exp(ik_{mx}x + ik_{mz}z), (l = 0, \pm 1) \quad (1)$$

$$H = k_0^{-1}\sum_m(-\hat{x}k_{mz} + \hat{z}k_{mx})E_m \exp(i\varphi_m)\exp(ik_{mx}x + ik_{mz}z), (l = 0, \pm 1) \quad (2)$$

$$S_z = <E \times H>_z = \mathrm{Re}(E_x H_y^* - E_y H_x^*), \quad (3)$$

Here, $k_{mx}^2 + k_{mz}^2 = k_0^2$, $k_{mx}=2m\pi/d$, $k_0=2\pi/\lambda$. $E_m$ and $\varphi_m$ are the real amplitude of the $m$th diffractive orders, respectively. In order to reconstruct the near field image, the phase and diffraction efficiency of each order should be calculated firstly. The rigorous couple wave analysis, a pure numerical method, may hide the physical insight into the diffraction process. Fortunately, the simplified modal method can offer us a vividly physical picture without complicated calculation. Thus, based on the simplified modal method, electric field of three diffractive orders can be expressed as [8]:

$$E_0 = \frac{1}{d}\int_{-d/2}^{d/2}[a_0 u_0(x)e^{ik_0 n_{0eff} h} + a_2 u_2(x)e^{ik_0 n_{2eff} h}]dx, \tag{4}$$

$$E_1 = \frac{1}{d}\int_{-d/2}^{d/2}[a_0 u_0(x)e^{ik_0 n_{0eff} h} + a_2 u_2(x)e^{ik_0 n_{2eff} h}]e^{-ik_x x}dx, \tag{5}$$

$$E_{-1} = \frac{1}{d}\int_{-d/2}^{d/2}[a_0 u_0(x)e^{ik_0 n_{0eff} h} + a_2 u_2(x)e^{ik_0 n_{2eff} h}]e^{ik_x x}dx, \tag{6}$$

The diffraction efficiency and the phase of $i$th order can be expressed as:

$$\eta_i = |E_i|^2, \tag{7}$$

$$\varphi_i = phase(E_i), \tag{8}$$

Now the electric field $E_0$ can also be expressed as:

$$E_0 = \frac{1}{d}\int_{-d/2}^{d/2} a_0 u_0(x)e^{ik_0 n_{0eff} h}dx + \frac{1}{d}\int_{-d/2}^{d/2} a_2 u_2(x)e^{ik_0 n_{2eff} h}dx,$$

$$= c_{00}e^{ik_0 n_{0eff} h} + c_{02}e^{ik_2 n_{2eff} h} \tag{9}$$

Where

$$c_{0i} = \frac{1}{d}\int_{-d/2}^{d/2} a_i u_i(x)dx, \tag{10}$$

And the electric field $E_1$ can also be expressed as:

$$E_1 = \frac{1}{d}\int_{-d/2}^{d/2} a_0 u_0(x)e^{ik_0 n_{0eff} h}e^{-i2\pi x/d}dx + \frac{1}{d}\int_{-d/2}^{d/2} a_2 u_2(x)e^{ik_0 n_{2eff} h}e^{-i2\pi x/d}dx,$$

$$= c_{10}e^{ik_0 n_{0eff} h} + c_{12}e^{ik_0 n_{2eff} h} \tag{11}$$

Where

$$c_{1i} = \frac{1}{d}\int_{-d/2}^{d/2} a_i u_i(x)\cos(2\pi x/d)dx, \tag{12}$$

Now we define two functions:

$$F_1(h) = c_{10}\cos(k_0 n_{0eff} h) + c_{12}\cos(k_2 n_{2eff} h), \tag{13}$$

$$F_2(h) = c_{10}\sin(k_0 n_{0eff} h) + c_{12}\sin(k_2 n_{2eff} h), \tag{14}$$

Therefore, the electric field $E_1$ can be expressed as:

$$E_1 = F_1(h) + iF_2(h), \quad (15)$$

$u_m(x)$ [9] is the electric field distribution of the $m$th grating mode. Because of even symmetry of incidence light, only two even grating modes can be excited, which determine the near-field image of Talbot effect. Obviously, the phase and diffraction efficiency of each order is the periodical function of variable $h$ and the period $T$ is:

$$T = \frac{2\pi}{k_0 n_{0eff} - k_0 n_{2eff}} \quad (16)$$

Now, we present that both the phases and diffraction efficiencies are periodically functions, which have the same period $T$. In order to test of the periodicity of phase and diffraction efficiency of each order, we calculate the phases and diffraction efficiencies of three orders versus the grating depth $h$. From the Figure 2, we can find that both phase and diffraction efficiency are the periodically functions.

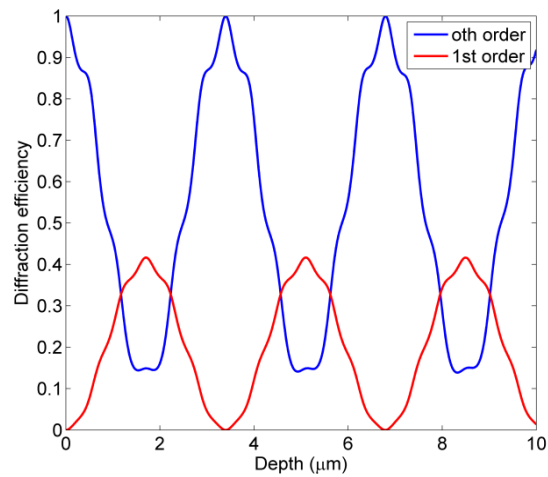

2(a)

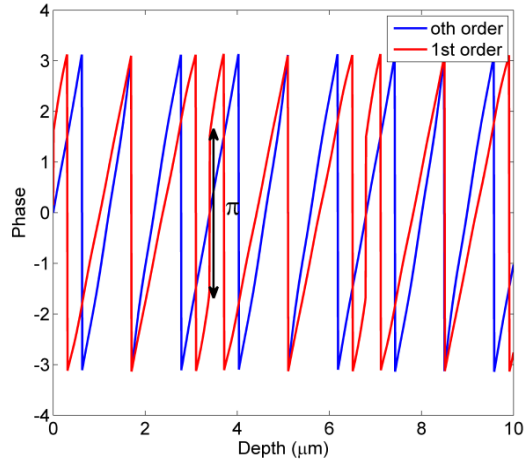

2(b)

Figure 2 The diffraction efficiencies (a) and phases (b) calculated by using the rigorous couple wave analysis. Grating parameters: duty cycle $f$=0.5, period=1.5$\lambda$.

In order to test the periodicity of the Talbot effect, the reconstructed images are shown in Fig. 3 for depth=1.7μm and depth=5.1μm, respectively. For simplicity, the time averaged Poynting vector $<S>=<E \times H>$ is set to be unit. The z-component time-averaged Pointing vector has almost the same near-field distribution for both depths. This identical phenomenon can be attributed to the approximate equal diffraction efficiencies and phases with corresponding diffractive orders, as shown in Table 1.

Table 1 The diffraction efficiencies and phases obtained by the RCWA for the depth=1.7μm and depth=5.1μm, respectively.

| Order | | -1$^{st}$ order | 0$^{th}$ order | 1$^{st}$ order |
|---|---|---|---|---|
| Diffraction efficiency (%) | Depth=1.7μm | 41.66 | 14.89 | 41.66 |
| | Depth=5.1μm | 41.67 | 14.88 | 41.67 |
| Phase (degree) | Depth=1.7μm | 179.9 | -179.8 | 179.9 |
| | Depth=5.1μm | 179.2 | -179.2 | 179.2 |

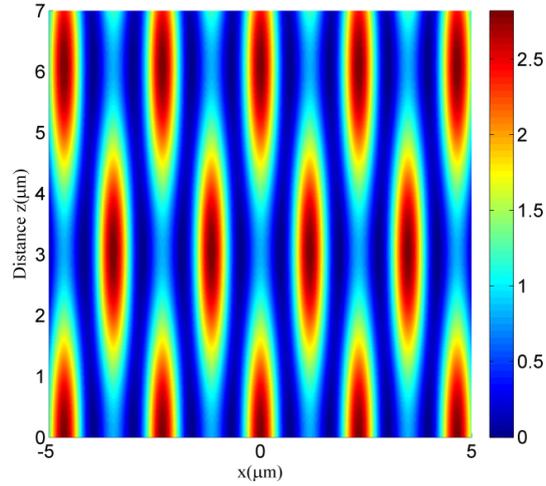

3(a)

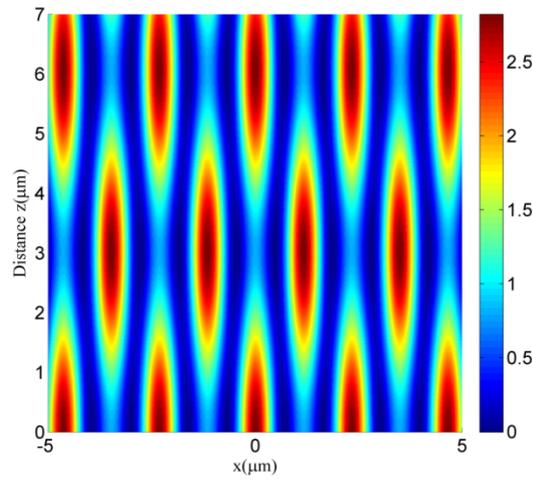

3(b)

Figure 3 The z-component of time-averaged Poynting vector <Sz>. (a) and (b) are the intensities for depth=1.7μm and depth=5.1μm directly obtained by RCWA.

Interestingly, π phase jump can occur at a special grating depth in Fig 2(b). This novelty phenomenon can be analyzed by the simplified modal method. Based on the Eq. (13-15), the phase of the first order can be expressed as:

$$\tan(\varphi_1) = \frac{c_{10}\sin(k_0 n_{0\mathit{eff}} h) + c_{12}\sin(k_2 n_{2\mathit{eff}} h)}{c_{10}\cos(k_0 n_{0\mathit{eff}} h) + c_{12}\cos(k_2 n_{2\mathit{eff}} h)}, \quad (17)$$

Figure 4 show profiles of functions $F_1$ and $F_2$ versus depth. The point of π phase jump is $h_0$, and the functions $F_1$ and $F_2$ are zero at this point. Thus we can use L'Hôspital's rule to analyze the phase jump at this point:

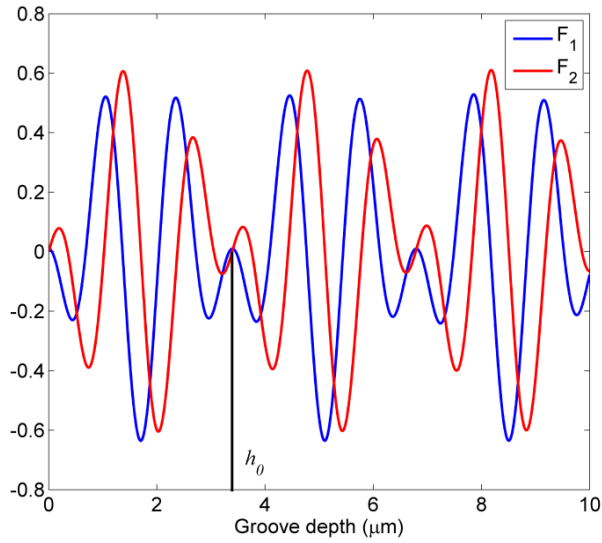

Figure 4 The profiles for functions $F_1$ and $F_2$, respectively.

The phase of the first order can be rewritten as:

$$\tan(\varphi_1(h_0)) = \lim_{x \to h_0} \frac{F_2(h)}{F_1(h)} = \frac{F_2'(h_0)}{F_1'(h_0)}, \quad (18)$$

The left derivative can be expressed as:

$$\tan(\varphi_1(h_0^+)) = \lim_{x \to h_0^+} \frac{F_2(h)}{F_1(h)} = \frac{F_2'(h_0^+) > 0}{F_1'(h_0^+) < 0} = -\infty, \quad (19)$$

And the right derivative can be expressed as:

$$\tan(\varphi_1(h_0^-)) = \lim_{x \to h_0^-} \frac{F_2(h)}{F_1(h)} = \frac{F_2'(h_0^-) > 0}{F_1'(h_0^-) > 0} = +\infty, \quad (20)$$

Thus we obtain the phase difference between both sides of $h_0$:

$$\varphi_1(h_0^+) - \varphi_1(h_0^-) = -\pi, \quad (21)$$

From above formulas, we can obtain that the phase jump at $h_0$ is indeed π. The physical reason of π phase jump is that the two even grating modes make the left derivative and right derivative of real part of the $E_1$ opposite in sign, one is positive infinity and another is negative infinity. Another implied condition of π phase jump is that $F_1$ and $F_2$ at the jump point are zero, which can lead to that $E_1$ is the zero. Therefore, the physical information of first order diffractive wave will be hidden in the near field image of Talbot effect. The jump point is determined by the grating mode index and mode coupling between the grating modes and diffractive orders. Also, the π phase jump can periodically occur with the same period $T$ mentioned above.

**Conclusion**

In this paper, the Talbot effect for a high density grating is interpreted by the simplified modal method, which can offer much vivid physical insight. The phases and diffraction efficiencies are periodicity for three orders and the period $T$ can be easily obtained by using the simplified modal method. By examination the phase of first order diffractive wave, the interesting phenomenon of π phase jump can occur since the two even grating modes make the left derivative and right derivative of real part of the $E_1$ opposite in sign. Because of π phase jump, the physical information of first order diffractive wave will be hidden in the near field Talbot effect image. A series of analytical formulas can be derived by the simplified modal method to depict the diffraction process. The simplified modal method is helpful for our understanding of the Talbot effect and π phase jump, which is also useful for applications in two- and three-dimentional nanostructure fabrication, polarization-controlled structure, structured illumination microscopy, etc..